\documentclass[letter,nofootinbib,12pt,superscriptaddress]{revtex4-2}
\usepackage[colorlinks,citecolor=blue]{hyperref}
\usepackage{amsmath}
\usepackage{mathrsfs}
\usepackage{bbm}
\usepackage{amsfonts}
\usepackage{amssymb}
\usepackage{latexsym}
\usepackage{graphicx}
\usepackage[english]{babel}
\usepackage{multirow}
\usepackage{float}
\usepackage{url}
\usepackage{slashed}
\usepackage[compat=1.1.0]{tikz-feynman}
\usepackage{orcidlink}

\begin{document}
\title{  Magnetic Monopoles and Exotic States in\\ $SU(4)_c \times SU(2)_L \times SU(2)_R$}

\author{Thomas W. Kephart}
\orcidlink{0000-0001-6414-9590}
\email{tom.kephart@gmail.com}
\affiliation{Department of Physics and Astronomy, Vanderbilt University, Nashville, TN 37235, USA}

\author{Qaisar Shafi}
\affiliation{Bartol Research Institute, Department of Physics and Astronomy, University of Delaware, Newark, DE 19716, USA}
\date{\today}

\begin{abstract}
\noindent
{\bf Abstract:} \\In the Pati-Salam gauge symmetry $SU(4)_c \times SU(2)_L \times SU(2)_R$ (4-2-2, for short), the observed quarks and leptons of each family reside in the bi- fundamental representations $(4,2,1)$ and $({\bar 4},1,2)$. There exist, however, the fundamental representations $(4,1,1)$, $(1,2,1)$ and $(1,1,2)$ and their hermitian conjugates, which show the presence, in principle, of yet to be discovered color triplets that carry electric charge $\pm{e/6}$, and color singlet particles with charges of $\pm{e/2}$. These Standard Model charges are in full accord with the fact that the 4-2-2 model predicts the presence of a topologically stable finite energy magnetic monopole that carries two quanta  of Dirac magnetic charge, i.e., $4 \pi/e$, as well as color magnetic charge that is screened beyond the quark confinement scale.
The 4-2-2 model therefore predicts the existence of exotic baryons, mesons and leptons that carry fractional ($\pm{e/2}$) electric charges. Since their origin lies in the fundamental representations of 4-2-2,  these exotic particles may turn out to be relatively light, in the TeV mass range or so. The 4-2-2 magnetic monopole mass  depends on the 4-2-2 symmetry breaking scale which may be as low as a few TeV. 
\end{abstract}

 \maketitle

\section{Introduction}  
 In this article we study some low energy consequences of the Pati-Salam gauge symmetry $SU(4)_c \times SU(2)_L \times SU(2)_R$ (4-2-2) \cite{Pati:1974yy}, which can be spontaneously broken to the Standard Model (SM) in a number of different ways. Along with a right-handed neutrino the observed quarks and leptons of each family reside in the bi-fundamental representations
$(4,2,1)$ and $({\bar 4},1,2)$ of 4-2-2. This feature is nicely explained if the 4-2-2 symmetry is embedded inside the grand unified symmetry group $SO(10)$
(more precisely $Spin (10)$). The decomposition of the 16 dimensional spinor representation of $SO(10)$ under 4-2-2 then yields the desired bi- fundamental representations.

However, if we do not assume that the 4-2-2 symmetry is embedded inside SO(10), we are led to consider the possibility that fields (bosons and/or fermions) in the fundamental representations (and their Hermitian conjugates) of 4-2-2
may exist in nature. Among other consequences, this means that the electric charge in 4-2-2 is quantized in units of $\pm{e/6}$, and the particles carrying this charge are also color triplets, hence exotic quarks, or e-quarks. Because of the $SU(4)_c$ symmetry there also appear color singlet leptons with electric charges of $\pm{e/2}$, i.e., e-leptons. These SM charges are compatible with the fact that the 4-2-2 model contains a topologically stable finite energy monopole that carries two quanta ($4\pi /e$) of Dirac magnetic charge \cite{Dirac:1931kp} as well as color magnetic field  \cite{Lazarides:1980cc}.

Recall that topologically stable finite energy monopoles arise from the spontaneous symmetry breaking of a gauge symmetry $G$ with manifest electric charge quantization to the SM. ‘t Hooft \cite{tHooft:1974kcl} and Polyakov \cite{Polyakov:1974ek} provided the first example of this phenomenon based on a toy model in which an SU(2) gauge symmetry is spontaneously broken with a scalar Higgs triplet to a $U(1)$ symmetry. Identifying the three gauge bosons in this toy $SU(2)$ model with $W^+$, $W^-$ and photon, and $U(1)_{em}$ as the unbroken symmetry, these authors showed the appearance of a topologically stable monopole in this model carrying two units ($4 \pi /e$) of Dirac magnetic charge and with a mass proportional to the symmetry breaking scale of the $SU(2)$ gauge symmetry.

The 4-2-2 monopole mass is proportional to the symmetry breaking scale of the underlying symmetry which may even lie in the low TeV range. As we increase the 4-2-2 breaking scale the monopole mass rises and can even approach the superheavy scale of order $10^{17}$ GeV, which is normally associated with the GUT monopole.

The exotic quarks in the $(4,1,1)+({\bar 4},1,1)$   representations of the 4-2-2 gauge symmetry give rise to exotic baryons and mesons carrying fractional electric charge ($\pm{e/2}$.)  The masses of these states can vary anywhere from the TeV region to the superheavy scale. Below we designate all exotically charges particles by appending an ``e'' prefix as above for e-leptons,  e-quarks, and likewise for e-baryons, e-mesons, etc. For a study of exotic states and magnetic monopoles in trinification models based on $SU(3)_c \times SU(3)_L \times
SU(3)_R$ (see Raut et al., \cite{Raut:2022ryj} and references therein.)

\section{$SU(4)_c \times SU(2)_L \times SU(2)_R$ monopole}
\label{4-2-2_monopole}

An elegant extension of the standard model is based on the gauge symmetry G = $SU(4)_c \times SU(2)_L \times SU(2)_R$  \cite{Pati:1974yy}.  The observed quarks and leptons of each family reside in the bi-fundamental representations $({4}, {2}, {1})$ and $(\overline{4}, {1}, {2})$ of G. The electric charge generator is  given by
\begin{equation}
    Q_{em} = \frac{(B - L)}{2}\;+\;\frac{T_{3R}}{2}\;+\;\frac{T_{3L}}{2},
    \label{qem}
\end{equation}
where $B$ and $L$ denote baryon and lepton numbers, and $T_{3L}$, $T_{3R}$ denote the third generators of $SU(2)_L$ and $SU(2)_R$ respectively. We normalize the $U(1)$ generators as in Ref.~\cite{Slansky:1981yr}, so that they have the minimal integer charges compatible with a period of $2 \pi$.
Note that G here is a global direct product of the three symmetry groups, and so the first homotopy group of G is trivial.
Among other things, this implies that electric charge is quantized in units of $e/6$. This can be seen by considering the fundamental representation $(4,1,1)$ of G whose components carry electric charges given by ($e/6, e/6, e/6, -e/2$). The first three components are $SU(3)_c$ color triplets, and the fourth component is a color singlet particle with electric charge $e/2$ [For a related discussion and additional references, see \cite{Raut:2022ryj,Kephart:2001ix,Kephart:2006zd,Kephart:2017esj}].
If the 4-2-2 symmetry is spontaneously broken to the SM at a few TeV, as discussed in \cite{Dolan:2020doe}, the predicted monopole acquires a mass in the multi-TeV range and may be accessible at an upgraded LHC or future colliders. Being highly relativistic such a monopole may be accessible in cosmic ray searches \cite{Kephart:1995bi,Kephart:2017esj,Frampton:2024shp}.  

\newpage
Next let us calculate the minimum Dirac magnetic charge the 4-2-2 monopole carries. 
\begin{figure}[h!]
\centering
      \includegraphics[width=0.8\textwidth,angle=0]{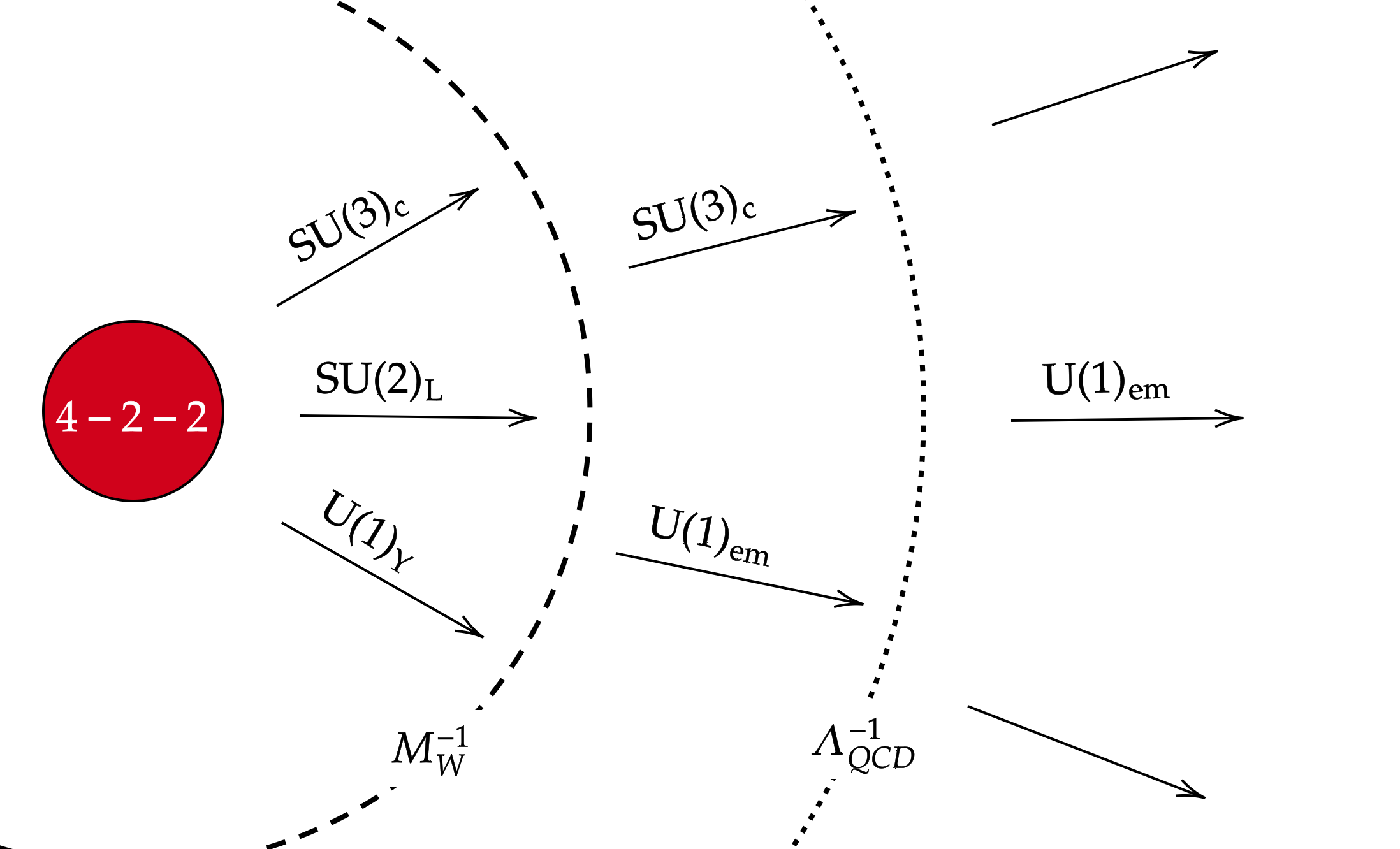}
        \caption{Schematic diagram showing the structure of the 4-2-2 monopole. The red core represents the region where the 4-2-2 symmetry is restored. Outside the  4-2-2  core and until \( M_W^{-1} \), the magnetic fluxes are associated with \( SU(3)_c \), \( SU(2)_L \), and \( U(1)_Y \). Beyond \( M_W^{-1} \) and until \( \Lambda_{\text{QCD}}^{-1} \), only \( SU(3)_c \)   and \( U(1)_{\text{em}} \)  fluxes are present. Beyond \( \Lambda_{\text{QCD}}^{-1} \), only the  electromagnetic flux remains. This monopole carries two units $(4\pi/e)$ of Dirac charge and its color magnetic flux is screened beyond $\Lambda^{-1}_{QCD}$.}
  \label{confinedcolor}
\end{figure}

\vspace{0.5cm}
\textbf{i)} We first consider the direct breaking of 4-2-2 to the SM. The SM hypercharges of the fundamental representation $(4,1,1)$ are given by
\begin{equation}
    Y = \left( \frac{1}{6}, \frac{1}{6}, \frac{1}{6}, -\frac{1}{2} \right). 
    \label{ygen}
\end{equation}
Note that in this case the hypercharge $Y$ coincides with the electric charge generator $Q_{em}$.
To identify the 4-2-2 monopole we note that a $4\pi$ rotation with $Y$ in Eq.~(\ref{ygen}) yields an identity element in the fourth component, but in the first three elements we end up with $e^{2i \pi/3}$, which lies in $SU(3)_c$. We therefore perform a $4\pi$ rotation along the color hypercharge generator
$diag (1/3, 1/3, -2/3, 0)$, which brings us back to the identity element.

We conclude that the 4-2-2 monopole carries two quanta $(4 \pi/e)$ of Dirac magnetic charge, as well as some color magnetic charge which is subsequently screened beyond the quark confinement radius (see Figure~\ref{confinedcolor}). 
Note that the $4 \pi$ rotation above corresponds to a $2 \pi$ rotation along the $SU(4)_c$ generator $B-L +
\frac{2}{3} T_c^8 = diag (1,1,-1,-1)$. This yields a closed loop in $SU(4)_c$, which is consistent with the fact that $SU(4)_c$ is simply connected.

\vspace{0.5cm}
\textbf{ii)} Another way to examine the 4-2-2 monopole is to consider the breaking of this symmetry to the SM via the the phenomenologically interesting subgroup $SU(3)_c \times SU(2)_L \times SU(2)_R \times U(1)_{B-L}$ \cite{Mohapatra:1974hk,Mohapatra:1974gc,Senjanovic:1975rk}. To find a stable monopole, we study the first homotopy group, 
\begin{equation}
  \pi_{1}\bigl(SU(3)_c \times SU(2)_L \times SU(2)_R \times U(1)_{B-L}\bigr).
\end{equation}
Acting on the quarks, a $2\pi$ rotation along the generator $B-L$  leads us to the center of $SU(3)_c$. Acting on the leptons it brings us to  the identity element. Next we preform a $2\pi$ rotation along $\frac{2}{3} T_c^8$, which  brings us back to the identity also in the quark sector. (Note that $U(1)_{B-L}$ intersects $SU(3)_c$ in its center.)
Furthermore, since the first homotopy groups of $SU(2)_{R}$ and $SU(2)_{L}$ are topologically trivial, we can include $2\pi$ rotations along the generators $T_{3R}$ and $T_{3L}$ without impacting the previous argument.

Put differently, a \(4\pi\) rotation along the generator
\begin{equation}
  \frac{\bigl(B - L\bigr)}{2}\;+\;\frac{T_{3R}}{2}\;+\;\frac{T_{3L}}{2}\;+\;\frac{T_{c}^{8}}{3}\ =Q_{em} +\frac{T_{c}^{8}}{3}\
\end{equation}
brings us back to the identity, which corresponds to a 4-2-2 monopole carrying two units $(4\pi/e)$ of Dirac magnetic charge. 
The presence of the color hypercharge generator shows that the monopole also carries color magnetic charge.

It is worth pointing out that the direct  breaking of SO(10) to the left-right symmetry group in Eq.(3) yields a monopole that carries only a single unit ($2\pi / e$) of Dirac magnetic charge \cite{Maji:2025yms}. This is intimately related to the fact that the embedding of left-right symmetry in $SO(10)$ yield new intersections between the topological spaces.

\vspace{0.5cm}

\textbf{iii.} Another interesting 4-2-2 symmetry breaking pattern is as follows:
\begin{equation}
\begin{split}
    SU(4)_c \times SU(2)_L \times SU(2)_R
    &\longrightarrow
SU(4)_c \times SU(2)_L \times U(1)_R \\
&\longrightarrow
SU(3)_c \times SU(2)_L \times U(1)_Y     
\end{split}
\label{422chainx}
\end{equation}
The first step in the breaking produces a monopole associated with a  $2\pi$ rotation along the generator $T_R^3$.
Since both $SU(4)_c$ and $SU(2)_L$ are simply connected spaces, we can include $2\pi$   rotations along their respective generators $X = B-L + 2/3 Y_c^8$ and $T_L^3$, without affecting the preceding topological argument. In other words, we have a monopole associated with a $4\pi$  rotation along the generator $Q_{em} + 1/3 Y_c^8$, where $Q_{em}$ is the generator of electromagnetic charge in 4-2-2, as shown in Eq.(\ref{422chainx}). Thus, the topologically stable finite energy 4-2-2 monopole carries two units $4\pi/e$  of Dirac magnetic charge as well as some color magnetic charge.

\vspace{0.5cm}

\begin{figure}
\centering
      \includegraphics[width=0.8\textwidth,angle=0]{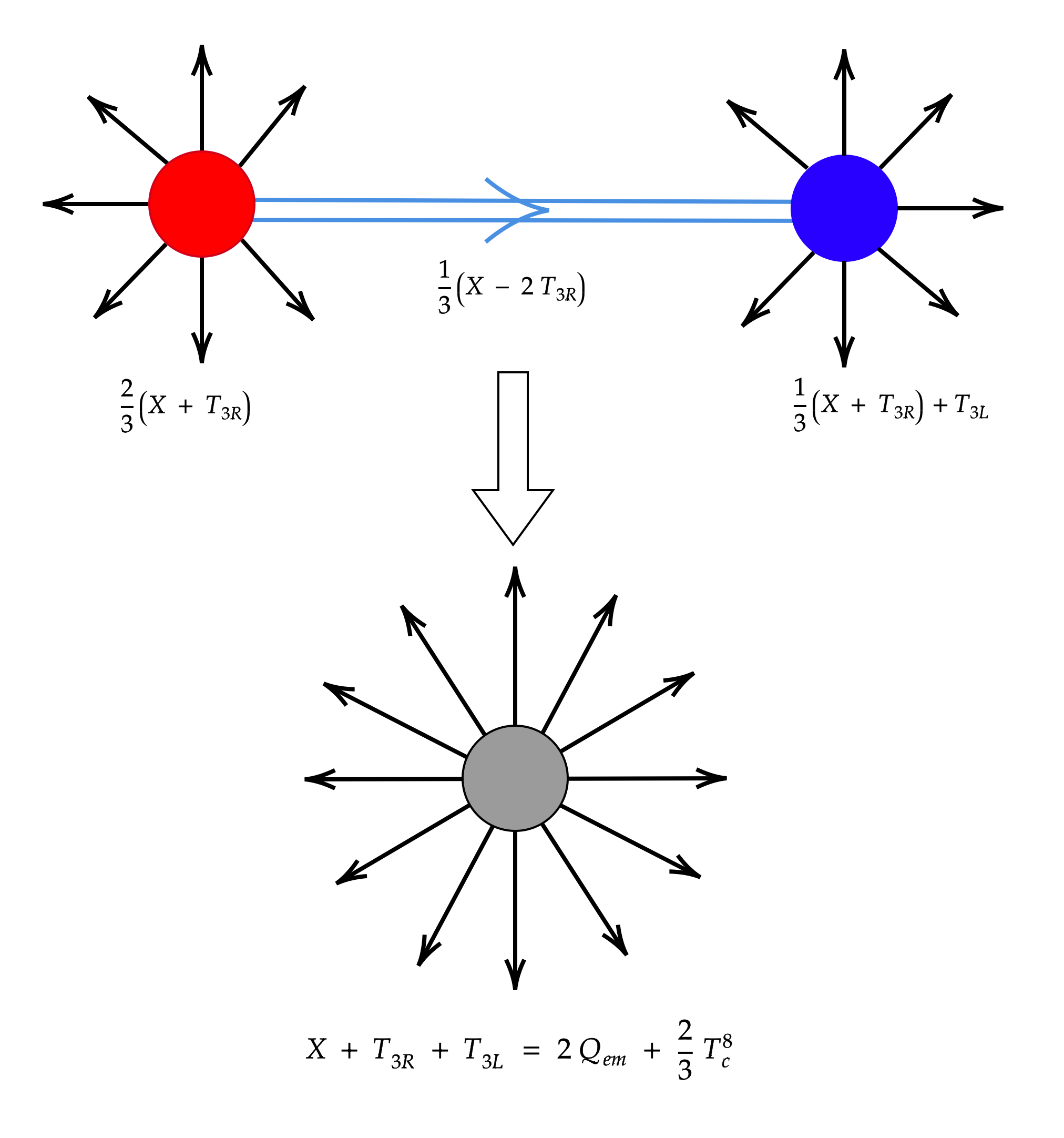}
        \caption{Emergence of the 4-2-2 magnetic monopole with two units of Dirac charge from the symmetry breaking $SU(4)_c \times SU(2)_L \times SU(2)_R \rightarrow SU(3)_c \times U(1)_{B-L} \times SU(2)_L \times U(1)_R \rightarrow SU(3)_c \times SU(2)_L \times U(1)_Y.$ This monopole also carries color magnetic charge. An $SU(4)_c$ (red) and an $SU(2)_R$ (blue) monopole are connected by a flux tube which pulls them together to form the topological stable 4-2-2 monopole. The magnetic flux along the tube and the Coulomb magnetic fluxes of the monopoles are indicated. Intermediate mass monopoles such as this one may survive inflation.}
  \label{redandblue}
\end{figure}

\textbf{iv.} Finally, let us consider the breaking of 4-2-2 to the SM via the  breaking chain:
\begin{equation}
\begin{split}
    SU(4)_c \times SU(2)_L \times SU(2)_R
    &\longrightarrow
SU(3)_c \times U(1)_{B-L} \times SU(2)_L \times U(1)_R\\
&\longrightarrow SU(3)_c \times SU(2)_L \times U(1)_Y
\end{split}
\label{422chain}
\end{equation}
The first breaking yields two kinds of magnetic monopoles which correspond to the breaking of $SU(4)_c$ to $SU(3)_c \times U(1)_{B-L}$, and $SU(2)_R$ to $U(1)_R$. They were coined red and blue monopoles respectively in Ref.~\cite{Lazarides:2019xai}. The breaking of $U(1)_{B-L} \times U(1)_R$ to $U(1)_Y$ generates a tube that connects a red with a blue monopole, and which  pulls the two monopoles together, to form the 4-2-2 monopole, as illustrated in Figure~\ref{redandblue}. For more details see Refs.~\cite{Lazarides:2019xai, Lazarides:2023iim}. 
Following Ref. \cite{Lazarides:2021tua}, we could refer to the red and blue monopoles as ‘magnetic quarks’ which carry Coulomb as well as confined magnetic flux and do not therefore exist as isolated states. Their merger yields the 4-2-2 monopole, namely a ‘magnetic hadron’ that carries a conserved magnetic charge.
It remains to be seen if the topological structure shown in Figure \ref{redandblue} can be found in condensed matter physics.

\section{Primordial (4-2-2) monopoles}
\label{primordial_monopole}
\subsection{Production}
In many scenarios magnetic monopoles are produced, often in abundance, in  early Universe phase transitions.
The number density  of monopoles emerging from
a phase transition is determined by the Kibble mechanism \cite{Kibble:1980mv},
where approximately one monopole
or anti-monopole is produced per correlated volume.
The resulting monopole number density today is \cite{Kolb:1990vq}
\begin{equation}
n_M \sim 10^{-19}\, (T_c/10^{11}{\rm GeV})^3 (l_H/\xi_c)^3\,{\rm cm}^{-3},
\label{density}
\end{equation}
where $\xi_c$ is the phase transition correlation length,
bounded from above by the horizon size $l_H$ at the time when the
region relaxes to the  broken--symmetry phase.
In a strongly first order transition,
the correlation length is much smaller than the horizon size.
In a second order or weakly first order phase transition,
the correlation length is comparable to the horizon size.
While monopoles could have been diluted by annihilation or inflation, they may still  exist in detectable numbers today
in some models.
Next we consider three scenarios according to which an observable number density of primordial monopole may be present in our galaxy.
The first example is based on the following symmetry breaking of SO(10), which yields intermediate scale 4-2-2 monopoles:
\begin{equation}
    SO(10) \longrightarrow SU(4)_c \times SU(2)_L \times SU(2)_R \longrightarrow SU(3)_c \times SU(2)_L \times U(1)_Y
\end{equation}
The first breaking yields the superheavy GUT monopole, while the second breaking yields a topologically stable intermediate mass 4-2-2 monopole.
According to the inflation model described in \cite{Maji:2022jzu}, the GUT monopole is inflated away, but the 4-2-2 monopole with mass comparable to the Hubble parameter during inflation experiences a limited number of e-foldings. The 4-2-2 monopole number density is estimated as follows:

After reheating, the monopole yield \(Y_{M}\) takes the form
\begin{equation}
Y_M \equiv \frac{n_M}{s}\simeq \frac{\frac{\xi_c^{-3}}{10} \exp(-3N_M) \left(\frac{\tau}{\tau_r}\right)^2} {\frac{2\pi^2}{45} g_* T_r^3},
\label{eq:YM-reheating}
\end{equation}
where $\xi_c \;=\; \min\;\bigl(H^{-1},\; m_{\mathrm{eff}}^{-1}\bigr),$ specifies the corrolation length at the time of production. The term in the denominator is the entropy density after reheating. The parameter \(N_{M}\) denotes the number of e-folds during which monopoles are produced, while \(\tau\), \(t_r\), \(g_*\), and \(T_r\) represent the inflation decay time, reheating time, effective number of relativistic degrees of freedom, and temperature at reheating, respectively. The two major effects that reduce the final monopole abundance are the exponential dilution from inflation, \(\exp(-3N_M)\), and the additional dilution from inflation oscillations, \(\left(\tau/t_r\right)^2\).
In light of the current experimental constraints, the MACRO collaboration sets an upper limit \(Y_{M} \lesssim 10^{-27}\). On the other hand, for
any feasible detection, one would typically require
\(Y_{M} \gtrsim 10^{-35}\).

Our second example comes from a recent paper \cite{Lazarides:2024niy} that shows how an observable number density of superheavy monopoles can arise from quantum mechanical tunneling in the early universe. This  scenario exploits inflation, metastable cosmic strings and the presence of red and blue monopoles as shown in Figure \ref{redandblue}.

In our final example \cite{Moursy:2024hll}, which can be extended to the 4-2-2 model, the monopole is associated with a waterfall field which experiences a limited number of $e$-foldings.

\section{Magnetic Monopoles as Cosmic Rays}

Magnetic monopoles (MMs) are natural candidates for ultra high energy cosmic rays (UHECRs). See \cite{Kephart:1995bi,Wick:2000yc} and references therein.
In this section we summarize the expected interactions of various types of monopoles with matter and the possibilities of their detection in cosmic ray observatories.

\subsection{Monopole Acceleration}
What makes magnetic monopoles such interesting  candidates for UHECRs is that all types of monopoles can be accelerated to energies above $\sim 10^{20}$ eV 
without invoking speculative mechanisms. All one needs are magnetic fields, which are abundant all over the Universe. A free monopole is accelerated along the field lines
to high energies. 

The kinetic energy gained by a magnetic monopole on traversing a magnetic
field along a  path is \cite{Kephart:1995bi}
\begin{equation}
E_K=g \int_{\rm{path}} \vec{B} \cdot\vec{dl}\,
\sim \sqrt{n}\,
\label{Ekin}\,g\,B\,\ell\,,
\end{equation}
where
\begin{equation}
g=e/2\alpha= 
 3.3\times 10^{-8} \,\, dynes/G
\label{charge}
\end{equation}
is the   magnetic charge,
$B$ is the magnetic field strength, $\ell$ is the
field's coherence length, and $\sqrt{n}$ is
a factor to roughly approximate the random--walk through the $n$ domains
of coherent fields traversed by the monopole path.
%

%%%%%%%%%%%%%%%%%
%\begin{figure}
%\includegraphics{MonoTable1.pdf.png}
%\caption[Monopole Energies.]
%{Estimated magnetic field strength and coherence length for some astrophysical environments, and the associated
%monopole energies for a single transit through the regions.}
%\label{MMenergies}
%\end{figure}
%%%%%%%%%%%%%%%%
%
\begin{table}
\begin{tabular}{|l|c|c|r|} \hline
  & $B/\mu$G & $\ell$/Mpc & $gB\ell$/eV  \\ \hline
normal galaxies & $3 \;{\rm to}\;10$ & $10^{-2}$ & $(0.5\;{\rm
to}\;1.7)\times 10^{21}$
      \\ \hline
starburst galaxies & $10 \; {\rm to}\; 50$ & $10^{-3}$ &
      $(1.7\;{\rm to}\;8)\times 10^{20}$ \\ \hline
AGN jets & $\sim 100$ & $10^{-4} \; {\rm to} \; 10^{-2}$ &
            $1.7\times (10^{20}\;{\rm to}\;10^{22})$   \\
\hline
galaxy clusters & $5\; {\rm to}\; 30$ & $10^{-4} \; {\rm to} \; 1$ &
     $3\times 10^{18}\;{\rm to}\; 5\times 10^{23}$   \\ \hline
Extragal. sheets & $ 0.1 \; {\rm to} \; 1.0$ & 1 to 30 &
     $1.7\times 10^{22}\;{\rm to}\;5\times 10^{23}$   \\ \hline
\end{tabular}
\caption{
Estimated magnetic field strength ${\bf B}$, coherence length $\ell$ and energies in eV for some astrophysical environments, and the associated
monopole energies for a single transit through the regions. For more details and references see \cite{Wick:2000yc}.}
\label{MMenergies} 
\end{table}

\subsubsection{Magnetic Field Regions}

Table \ref{MMenergies} list the typical energy gained by a MM in traversing various regions of the Universe.
In the table we collect the cosmic magnetic fields 
and their
coherence lengths, inferred from observations of synchrotron radiation,
Faraday rotation, and  models.
The typical  monopole kinetic energies that result
are all above the Greisen–Zatsepin–Kuzmin (GZK) bound \cite{Greisen:1966jv,Zatsepin:1966jv}.

\subsection{Summary of UHECR Observations}
 The GHZ bound limits the distance ultra high energy cosmic ray (UHECR) protons can travel due to interactions with cosmic background photons.
\begin{equation}
 p+\gamma\rightarrow \Delta^* \rightarrow p+\pi \rightarrow p+ ...
\label{GZK}
\end{equation}
The threshold for this process is about $5\times 10^{19}$ eV, leading to a pileup of cosmic ray protons at this energy.
But, super GZK energy events have been seen. This implies the initial protons were accelerated nearby, since the mean free path
of a super GZK proton is a few Mpc, or that some of the observed events are not protons. 

The highest energy cosmic ray ever seen is the HiRes (The High Resolution Fly's Eye Ultra 
High Energy Cosmic Ray Observatory) event at energy
$$3.12\times10^{20}\,\, eV.$$
There are now several dozen events above the GZK cutoff. Remarkably, some come from the direction of the local void. As an example, the recently observed
Amaterasu event of energy $$2.40\times10^{20}\,\, eV,$$ tied (within error bars) for second with two other events from the AGASSA observatory, is one such event.
The local void is about 45 Mpc across with no known sources of ultra high energy protons and so it is about 10 proton mean free paths  (mfps) across, making the idea that the Amaterasu event 
may be a magnetic monopole primary seem not unlikely.

There are a number of other UHECR events of transGZK energies that appear to be coming reom the direction of the local void. For a discussion see \cite{KP}.

\subsection{Detection}
While all MMs will accelerate to the same energy in a magnetic field, up to a factor of the charge of the monopole, where its charge is some
integer $n$ times the Dirac charge, the behavior of the monopoles when interacting with matter  can be very different, depending on their characteristics, 
including mass, charge, and their
interaction strengths--strong, weak and/or electromagnetic.
Even though there are a large number of
possible monopole types, they should each have their
own signature. Some could be best seen in air showers,
others may need ICE CUBE or something else.
But cosmic ray detectors have 
the best and perhaps only realistic chance 
of seeing them.
However, as low mass monopoles could potentially be see in accelerators, we refer the reader to
\cite{ahmed2024magneticmonopolephenomenologyfuture}
for a discussion of magnetic monopole phenomenology at future hadron colliders.

\subsubsection{Detectors/Experiments/Observatories}
Pioneering UHECR observations were carried out at
Volcano Ranch, Haverah Park, SUGAR, Yakutsk, and Akeno arrays  which led to 
new more ambitious observatories with  improved detection techniques by the AGASA, 
Fly’s Eye, HiRes, Auger, Ice Cube and Telescope Array experiments. For reviews see \cite{Nagano:2000ve,Dawson:2017rsp}.

\subsubsection{Magnetic Monopole Cosmic Ray Shower Development}
\noindent
Monopole CR shower development depends 
on the monopole mass and structure.
The showers from monopoles that only have electromagnetic interactions are slow to develop 
(They keep most of their energy after a collision, so these monopoles remain at high energy and allow the shower to be are continuously initiated.)
By contrast, hadronic monopole showers can develop quickly 
and can look similar to proton or nucleus showers. The 4-2-2 monopole, we recall, carries color magnetic charge.

\subsubsection{Detection of Monopoles via Electromagnetic Interactions}
\noindent
Monopole electromagnetic energy losses are dominated by the following three processes \cite{Wick:2000yc}:\\
(i) Collisional Energy Loss; (ii) Pair Production; (iii) Monopole Photonuclear Interaction. \\
At low $\gamma$ the monopole energy loss is dominated by collisional loss. 
 This persists until about $\gamma= 10^4$ where pair production takes over. 
 This continues until approximately $\gamma= 10^6$ where photo-nuclear processes take over and dominate asymptotically in $\gamma$. 
 Other processes like Monopole Bremsstrahlung are always subdominant.

\noindent
Note that the polarization of Cherenkov radiation from monopoles is rotated 90 degrees from that of protons.
This provides the potential to identify monopoles early during
shower development.\\

\subsubsection{Ice Cube} Recently Ice Cube \cite{IceCube:2021eye} has set new limits on relativistic magnetic monopoles with $0.5 \le \gamma \le 0.995$. This improves on their previous bound \cite{Aartsen_2016},  and agrees with bounds from the ANTARES experiment \cite{ANTARES2017}.

 \subsubsection{ Detection of Hadronic Monopoles}
\noindent
The constituents of hadronic (baryonic or mesonic) monopoles  carry  magnetic charge and are attached via  magnetic flux tubes.
The red-blue mesonic case discussed above and shown in Figure \ref{redandblue} is an example.
Another example, not studied in this work, is  the topologically stable triply charged baryonic monopole  \cite{Lazarides:2021tua} from the symmetry breaking $G \rightarrow SU(3)_c \times SU(2)_L \times U(1)_{Y_L} \times U(1)_{Y_R} \times U(1)_R \rightarrow SU(3)_c \times SU(2)_L \times U(1)_Y \rightarrow SU(3)_c \times U(1)_{em}$, where the constituents are an $SU(2)_R$  monopole connected by a flux tube to an $SU(3)_L$ monopole which, in turn, is connected to an $SU(3)_R$  monopole by a superconducting flux tube. The constituent monopoles are pulled together to form the stable triply charged monopole. 

The initial energy of an UHECR is $E_{cr}\sim 10^{20} $ eV or about $ 10^{8}$ TeV. We assume that the symmetry breaking where a monopole appears is around or just above the electroweak scale, $\Lambda_{SB}\sim 1 $ TeV. However, that monopole has mass $M\sim \Lambda_{SB}/\alpha$, and hence is about two orders of magnitude heavier than $\Lambda_{SB}$. The center of mass energy of an UHECR monopole colliding with an atmospheric proton is 
 $E_{cm}=\sqrt{m_p^2+M^2 +2m_pE_{cr}}\sim 10^3$ TeV.
 The magnetic flux tubes can only break by producing a $M {\bar M}$ pair, and the flux  tube carries energy $\Lambda_{SB}$ per unit length. Hence the tube can stretch by a  factor of 
 $\Lambda_{SB}/\alpha$ before breaking.   
 This leads to a cross section that grows by a factor of $\sim 100$ after the first collision, assuming the flux tube has a similar cross section per unit length to the monopole itself.  Hence the excited monopole-flux tube system's cross section approaches that of a typical QCD hadronic cross section.  We conclude that the excitation of the initial highly relativistic ($\gamma \sim 10^7$ ) monopole of energy $E_{cr}$ and mass $M\sim 10$ TeV  can emulate an ultra high energy proton shower.

\section{Free Fractional Electric Charge}

 Monopoles and fractionally-charged color singlets go hand in hand in the class of models that include the 4-2-2 model discussed here.
 If the minimally charged monopole has multiple Dirac charge, then this  implies a fractionally-charged color singlets must be in the theory due to the relation $\frac{eg}{4\pi}=\frac{n}{2}.$
 Fractional electric charge has been of interest since Fairbanks et al. \cite{PhysRevLett.38.1011} redid the Millikan oil drop experiment of the early $20^{th}$ century. In \cite{PhysRevLett.38.1011} a superconducting magnetic levitation experiment was used. Two of the levitated niobium balls  were found to have residual charges of $(+0.337 \pm 0.009)$e  and $(-0.331 \pm 0.070)$e. 

If free fractional charges exist,  they could bind to form atoms. Suppose we have a massive $-\frac{1}{3}e$  charge particle \cite{Goldberg:1981jt,Frampton:1982gc} which binds to a proton and forms a hybrid hydrogen atom. The spectral lines in this case resemble hydrogen, but are shifted due to the potential's change of charge factor from $e^2$ to $e^2/3.$

Likewise, a free $+e/2 $ charged particle, as predicted by the 4-2-2 model, could bind directly to an electron.
Again the solution to the Schrodinger equation should be similar
to that for the hydrogen atom where $e^2  \rightarrow  e^2/2$ everywhere.
Since the energy levels for hydrogen are
$E_n = -(me^4)/(2 \hbar^2 n^2)$,
for this new exotic atom we get
$E_n = -(m e^4)/(8 \hbar^2 n^2)$.
 The spectrum is the same as hydrogen, except it is scaled
by a factor of 4. (It looks like  it has been red shifted.)
So one could search for e-hydrogen by looking for weak shifted hydrogen like lines
in stellar atmospheres or gas clouds.

\subsubsection{Current Fractional Charge Searches}

A variety of experiments are currently searching for magnetic monopoles and related  fractionally charged particles \cite{MoEDAL:2024wbc,ATLAS:2024kyv, ATLAS:2023esy, Wulz:2021jsu,IceCube:2021eye,Frank:2024sfv, Panda:2024vhc, NOvA:2020qpg}.
There is an extensive list of bifundamental models with fractional charged color singlets and multi-charged monopoles  yet to be fully explored, however these models are beyond the scope of this work \cite{Sheridan:2022qku}.

\subsubsection{The CMS Search for  Fractional Charge }
 Recently, the CMS Collaboration at CERN
 has searched for fractionally charged particles via a Drell-Yan-like production mode with $q<e$ in proton-proton collisions at $\sqrt{s} = 13$ TeV \cite{cmscollaboration2024searchfractionallychargedparticles}. They find the most stringent limits to date 
and exclude masses up to 640 GeV and charges as low as $e/3$ at the 95\% confidence level.  
 We emphasize that the lightest exotic charged particle must be stable since charge is conserved.

 \subsubsection{CUORE and the Search for  Fractional Charge }
 The Cryogenic Underground Observatory for Rare Events (CUORE) at the Laboratori Nazionali del Gran Sasso (LNGS),
 is a cryogenic calorimetric experiments able to search for fractionally-charged particles \cite{CUORE:2002myo}. They have recently placed a bound on the flux $\Phi$ of e-particles  in the range $e/24$ to $e/2$ at $\Phi < 6.9 \times 10^{-12} cm^{-2} s^{-1} Sr^{-1}$ (90\% C.L.)  \cite{PhysRevLett.133.241801}.
  
\section{Summary}
\label{summary}
 Relatively light monopoles in models based on the gauge symmetry $SU(4)_c \times SU(2)_L \times SU(2)_R$  should be accessible at an upgraded LHC or future colliders. This model also predicts e-particles, the existence of particles in exotic color singlet states that carry fractional electric charge. Heavier and intermediate mass monopoles can be searched for in cosmic ray experiments as well as with  neutrino detectors. Progress has recently been made in understanding how the superheavy  monopole can survive primordial inflation and which may be present in our galaxy at an observable level.
 For a discussion of magnetic monopole phenomenology at future hadron colliders see \cite{ahmed2024magneticmonopolephenomenologyfuture}.

\begin{acknowledgments}
\noindent
We thank Dr. Rinku Maji for helpful discussions.
Q.S. thanks Dr. Amit Tiwari for discussion and collaboration.
\end{acknowledgments}

% Bibliography

%% [A] Recommended: using JHEP.bst file
%\bibliographystyle{JHEP}
\bibliography{topological.bib}
%\bibliography{CosmicRay.bib}
%% or
%% [B] Manual formatting (see below)
%% (i) We suggest to always provide author, title and journal data or doi:
%% in short all the informations that clearly identify a document.
%% (ii) please avoid comments such as "For a review'', "For some examples",
%% "and references therein" or move them in the text. In general, please leave only references in the bibliography and move all
%% accessory text in footnotes.
%% (iii) Also, please have only one work for each \bibitem.

\end{document}